\documentclass[a4paper,12pt,colorlinks,urlcolor=black,linkcolor=black,citecolor=black,
]{article}
\usepackage{amsmath,amssymb}
\usepackage[dvips]{color}
\usepackage{cite}
\usepackage{hangcaption}
\usepackage{amsmath,amssymb}
\usepackage[dvips]{color}
\usepackage{cite}
\usepackage{eepic}
\usepackage{epic}
\usepackage{hangcaption}
\usepackage{amsmath}
\usepackage{amssymb}
\usepackage{amscd}
\usepackage{amsthm}
\usepackage{multicol}
\usepackage[dvipdfmx,%
 bookmarks=true,%
 bookmarksnumbered=true,%
 colorlinks=true,%
 setpagesize=false,%
 pdfkeywords={TeX; dvipdfmx; hyperref; color;}]{hyperref}
\def\cases{\left\{\begin{array}{ll}}
\def\endcases{\end{array}\right.}

\begin{document}
\setcounter{page}{1}
\vskip1.5cm
\begin{center}
{\Large \bf 
The realization of the wave function collapse in the linguistic interpretation of quantum mechanics
}
\vskip0.5cm
{\rm
\large
Shiro Ishikawa
}
\\
\vskip0.2cm
\rm
\it
Department of Mathematics, Faculty of Science and Technology,
Keio University,
\\ 
3-14-1, Hiyoshi, Kouhoku-ku Yokohama, Japan.
E-mail:
ishikawa@math.keio.ac.jp
\end{center}
\par
\rm
\vskip0.3cm
\par
\noindent
{\bf Abstract}
\normalsize
\vskip0.5cm
\par
\noindent
Recently I proposed the linguistic interpretation of quantum mechanics, which is characterized as the linguistic turn of the Copenhagen interpretation of quantum mechanics. This turn from physics to language does not only extend quantum theory to classical theory but also yield the quantum mechanical world view. Although the wave function collapse is prohibited in the linguistic interpretation, in this paper I show that the phenomenon like wave function collapse can be realized in the linguistic interpretation. And furthermore, I propose the justification of the von Neumann-L\"uders projection postulate. After all, I conclude that the wave function collapse should not be adopted in the Copenhagen interpretation.
\par
\noindent
\vskip0.5cm
\par
\noindent
Key phrases:  Copenhagen interpretation, Wave function collapse, von Neumann-L\"{u}ders projection postulate

\par

\def\Cal{\cal}
\def\bigstimes{\text{\large $\: \boxtimes \,$}}

\par
\noindent

\vskip0.2cm
\par
\noindent
\par
\noindent
\section{\large
Preparations
}
Recently in \cite{KeioPress}-\cite{KSTS}, I proposed measurement theory (i.e., quantum language, or the linguistic interpretation of quantum mechanics), which is characterized as the linguistic turn of the Copenhagen interpretation of quantum mechanics.
This turn from physics to language does not only extend quantum theory to classical theory but also yield the quantum mechanical world view. 
The linguistic interpretation says that 
\begin{itemize}
\item[(A)]\it
{\lq\lq}Only one measurement is permitted", 
\rm
and thus, 
we are not concerned with anything after measurement
since it 
can not be measured any longer.
Also,
the Heisenberg picture should be adopted,
that is,
the Schr\"{o}dinger picture
should be prohibited.
( For details, see \cite{JQIS1, JQIS2, KSTS}. )
\end{itemize}
Therefore, the wave function collapse is meaningless in the linguistic interpretation.
In this sense, the linguistic interpretation and the Copenhagen interpretation are different.
\par
Although my idea proposed in this paper was discovered in the investigation of quantum language, it may be understood 
without the knowledge of quantum language. Hence, the readers are not required to have the usual knowledge of quantum language, but that of quantum mechanics.

\par
\noindent
\subsection{\normalsize
Hilbert space
}

\rm
\par
\par
\noindent

According to ref.\cite{Neum},
we briefly introduce the mathematical formulation of quantum mechanics as follows.
\par
\par
\rm
\par
Consider an operator algebra $B(H)$
(i.e.,
an operator algebra
composed of all
bounded linear operators
on a Hilbert space $H$
with the norm
$\|F\|_{B(H)}=\sup_{\|u\|_H = 1} \|Fu\|_H$
), in which quantum mechanics is formulated.
Define 
$Tr (H)$, the trace class,
by $ Tr (H)= B(H)_*$ (i.e., pre-dual space).
For any
$u, v \in H$, 
define
$| u \rangle \langle v | \in B(H)$
such that
\begin{align}
(| u \rangle \langle v |)w = \langle v , w \rangle u
\quad (\forall w \in H) .
\label{eq1}
\end{align}
The trace map ${\rm tr}: Tr(H) \to {\mathbb C}$(= the complex field) is defined by
\begin{itemize}
\item[(B)]
$\qquad \qquad
{\rm tr}(T) = \sum_{k=1}^\infty \langle e_k , T e_k \rangle 
\quad (\forall T \in  Tr(H) )
$
\end{itemize}
where it does not depend on the choice of the complete orthonormal system
$\{ e_k \}_{k=1}^\infty $.
The mixed state space $Tr_{+1}(H)$ is defined by 
$\{ \rho \in Tr(H) \; | \; \rho \ge 0, \;\;\;  \mbox{tr}(\rho)=1 \}$.

\par
\noindent
\subsection{\normalsize
Observables, state, Markov operator
}
We define the observable ${\mathsf O}=(X, {\mathcal F}, F )$ in $B(H)$ 
(or, POVM, {\it cf} \cite{Davi}) such that\begin{itemize}
\item[(C$_1$)]
$X$ is set, ${\mathcal F}$ ($\subseteq 2^X$: the power set of $X$ ) is a $\sigma$-field.
\item[(C$_2$)]
$F: {\mathcal F} \to B(H)$ is a map such that $0=F(\emptyset ) \le F( \Xi ) \le F(X) =I$ (= the identity)
$(\forall \Xi \in {\mathcal F } )$,
\item[(C$_3$)]
for any countable decomposition
$\{\Xi_1, \Xi_2, \ldots, \Xi_n, \ldots\}$
of
$\Xi$
$\Big($
i.e.,
$\Xi = \bigcup\limits_{n=1}^\infty \Xi_n $, 
$\Xi_n \in {\cal F}, (n = 1, 2, \ldots)$,
$\Xi_m \cap \Xi_n =\emptyset 
\;\;
(m \not= n )$
$\Big)$, it holds that
\begin{align*}
\langle u , F(\Xi ) u \rangle = 
\lim_{n \to \infty }  \sum_{k=1}^n 
\langle u , F(\Xi_k ) u \rangle
\qquad (\forall  u \in H)
\tag{2}
\end{align*}
\end{itemize}
Also, a pure state is represented by $\rho = |u \rangle \langle u |$
( where $u \in H$,
$\|u \|=1 $
).
\par
Let $H_1$ and $H_2$ be Hilbert spaces.
A continuous linear operator
$\Phi : B(H_2) \to B(H_1)$ is said to be a Markov operator,
if the pre-dual operator $\Phi_* : Tr(H_1) \to Tr(H_2)$
satisfies that 
$\Phi_* ( Tr_{+1} (H_1) )
\subseteq Tr_{+1} (H_2)$.

\par
\noindent
\subsection{\normalsize
Axioms}
\par
A measurement of an observable
${\mathsf O}=(X, {\mathcal F}, F )$ 
for a state $\rho (= |\rho \rangle \langle u | )$ is denoted by
${\mathsf{M}}_{B(H)} ({\mathsf{O}}{\; :=} (X, {\cal F}, F),$
$ S_{[\rho]})$.
\par
Now we introduce two axioms as follows.

\par
\noindent
\bf
Axiom 1
\rm
[
Measurement
].
\it
The probability that a measured value $x$
$( \in X)$ obtained by the measurement 
${\mathsf{M}}_{B(H)} ({\mathsf{O}}{\; :=} (X, {\cal F}, F),$
$ S_{[\rho]})$
belongs to a set 
$\Xi (\in {\cal F})$ is given by
$$
\rho( F(\Xi) )
\Big(=
\mbox{\rm tr}(\rho F(\Xi)) =
\langle u, F(\Xi) u \rangle 
\Big)
$$
\rm
\par

\par
\rm
Axiom 2 is presented
as follows:
\rm
\par
\noindent
\bf
Axiom 2
\rm
[Causality].
\it
Let $t_1 \le t_2$.
The causality is represented by
a Markov operator 
$\Phi_{t_1,t_2}{}: $
$B(H_{t_2}) \to B(H_{t_1})$.

\rm

\par
\noindent
\section{\large
The wave function collapse
}
\par
\noindent
\subsection{\normalsize
The von Neumann-L\"{u}ders projection postulate
in the Copenhagen interpretation
}
Let
$H$ be a Hilbert space.
Let
${\mathbb P}=[P_k ]_{k=1}^\infty$
be a spectral decomposition in $B(H)$,
that is,
$P_k ( \in B(H) )$ is a projection
($\forall k =1,2,...$)
such that
\begin{align*}
\sum_{k=1}^\infty \langle u , P_k u \rangle  = \| u \|^2
\quad
(\forall u \in H )
\end{align*}
Put ${\mathbb N}=\{1,2,...\}$.
Define the observable
${\mathsf O}_P =( {\mathbb N},2^{\mathbb N}, P)$
in $B(H)$
such that
\begin{align*}
P( \{k \} ) = P_k
\qquad (\forall k =1,2,... )
\tag{3}
\end{align*}
Axiom 1 says:
\begin{itemize}
\item[(D$_1$)]
The probability that a measured value $n$
$( \in {\mathbb N})$ is obtained by a measurement 
${\mathsf{M}}_{B(H)} ({\mathsf{O}}_P$
${\; :=}( {\mathbb N},2^{\mathbb N}, P),$
$ S_{[\rho]})$
is given by
$$
\mbox{tr}( \rho P_n ) 
(= \langle u , P_n u \rangle ),
\quad (\mbox{ where } \rho= |u \rangle \langle u | )
$$
\end{itemize}
Also, the von Neumann-L\"{u}ders projection postulate
( in the Copenhagen interpretation, {\it cf.} \cite{{Luders}}) says:
\begin{itemize}
\item[(D$_2$)]
When a measured value $n$
$( \in {\mathbb N})$ is obtained by the measurement 
${\mathsf{M}}_{B(H)} ({\mathsf{O}}_P{\; :=}( {\mathbb N},2^{\mathbb N}, P),$
$ S_{[\rho]})$,
the state $\rho_a$ after the measurement is given by
\begin{align*}
\rho_a
=\frac{P_n |u \rangle \langle  u |P_n }{\| P_n u \|^2}
\Big(
=\frac{ |P_nu \rangle \langle P_n u | }{\| P_n u \|^2}
\Big)
\tag{4}
\end{align*}
And furthermore, when a measurement
${\mathsf{M}}_{B(H)} ({\mathsf{O}}_F{\; :=}(X,{\mathcal F}, F),$
$ S_{[\rho_a]})$ is taken, the probability that a measured value belongs to
$\Xi ( \in {\mathcal F} )$ is given by
\begin{align*}
\mbox{tr} ( \rho_a F(\Xi ))
\Big(=
\langle \frac{P_n u}{\| P_n u \| }, F(\Xi)
\frac{P_n u}{\| P_n u \|} \rangle
\Big)
\tag{5}
\end{align*}
\end{itemize}
Note that the von Neumann-L\"{u}ders projection postulate (D$_2$) is not adopted in our situation since the linguistic interpretation  (A) says that
the state after a measurement is meaningless.

\par
\noindent
\subsection{\normalsize
The von Neumann-L\"{u}ders projection postulate
in the linguistic interpretation
}

Consider a Hilbert space $H$ and a tensor Hilbert space $K \otimes H $.
Let ${\mathbb P}=[P_k ]_{k=1}^\infty$ be a spectral decomposition in 
$B(H)$,
and let $\{ e_k \}_{k=1}^\infty$ be a complete orthonormal system in
a Hilbert space $K$.
Define the pre-dual Markov operator
$\Psi_*: Tr(H) \to Tr(K \otimes H)$ 
by, for any $u \in H$,
\begin{align*}
\Psi_* (|u \rangle \langle u |) = \sum_{k=1}^\infty |e_k \otimes P_k u \rangle \langle e_k \otimes P_k u |
\tag{6}
\end{align*}
or
\begin{align*}
\Psi_* (|u \rangle \langle u |) =  |\sum_{k=1}^\infty(e_k \otimes P_k u )\rangle \langle \sum_{k=1}^\infty( e_k \otimes P_k u) |
\tag{7}
\end{align*}
Thus the Markov operator $\Psi: B(K \otimes H) \to B(H)$
is defined by
$\Psi = (\Psi_*)^*$.
\par
\noindent
Define the observable ${\mathsf O}_G = ( {\mathbb N}, 2^{\mathbb N}, G)$ in $B(K)$
such that
$$
G( \{k\} ) = |e_k \rangle \langle e_k |
\qquad
(k \in {\mathbb N} = \{ 1,2,... \})
$$
Let ${\mathsf O}_F = (X, {\mathcal F}, F)$
be arbitrary observable in $B(H)$.
Thus, we have the tensor observable
${\mathsf O}_G \otimes {\mathsf O}_F$
$=$
$({\mathbb N}  \times X, 2^{\mathbb N} \boxtimes {\mathcal F}, G \otimes F )$
in 
$B(K \otimes H)$,
where
$2^{\mathbb N} \boxtimes {\mathcal F}$
is the product $\sigma$-field.
\par
Fix a pure state $\rho = |u \rangle \langle u |$
$( u \in H , \| u \|_H = 1 )$.
Consider a measurement
${\mathsf M}_{B(H)} (\Psi( {\mathsf O}_G \otimes {\mathsf O}_F), S_{[\rho ]} )$.
Then, Axiom 1 says that
\begin{itemize}
\item[(E)]
the probability that a measured value
$( k , x )$ obtained by the measurement
${\mathsf M}_{B(H)}(\Psi ( {\mathsf O}_G \otimes {\mathsf O}_F), S_{[\rho ]} )$
belongs to $\{n\} \times \Xi $
is given by
\end{itemize}
\begin{align*}
&
\mbox{tr}[ (|u \rangle \langle u |) \Psi( G(\{n \}) \otimes F(\Xi))] 
=
\mbox{tr}[
(\Psi_* (|u \rangle \langle u |))
( G(\{n \}) \otimes F(\Xi))] 
\\
= 
&
\mbox{tr}
[(
\sum_{k=1}^\infty |e_k \otimes P_k u \rangle \langle e_k \otimes P_k u |
)
( 
| e_n \rangle \langle e_n |  \otimes F(\Xi))]
=
\langle P_n u , F(\Xi ) P_n u \rangle
\quad
(\forall \Xi \in {\mathcal F} )
\end{align*}
( In a similar way, the same result is easily obtained in the case of (7)).
\par
\noindent
Thus, we see:
\begin{itemize}
\item[(F$_1$)]
if $\Xi=X$, then we see:
\begin{align*}
&
\mbox{tr}[ (|u \rangle \langle u |) \Psi( G(\{n \}) \otimes F(X))] 
=
\langle u, P_n u \rangle 
\end{align*}
\item[(F$_2$)]
when a measured value $(k , x)$ belongs to $ \{ n \} \times X $,
the conditional probability such that $x \in \Xi$
is given by
\begin{align*}
\langle \frac{P_n u}{\| P_n u \| }, F(\Xi)
\frac{P_n u}{\| P_n u \|} \rangle
\quad (\forall \Xi \in {\mathcal F } )
\tag{8}
\end{align*}
\end{itemize}
This is a direct consequence of Axioms 1 and 2.
\par
Considering the correspondence: $\mbox{(D)} \Leftrightarrow \mbox{(F)}$, that is,
$$
{\mathsf{M}}_{B(H)} ({\mathsf{O}}_P,
S_{[\rho]})
\Leftrightarrow
{\mathsf M}_{B(H)}(\Psi ( {\mathsf O}_G \otimes {\mathsf O}_F), S_{[\rho ]} ),
\quad
\mbox{(D$_1$)} \Leftrightarrow \mbox{(F$_1$)},
\quad
\mbox{(D$_2$)} \Leftrightarrow \mbox{(F$_2$)}
$$
there is a reason to consider that
the true meaning of the (5) is just the (8).

\par
\noindent
\section{\large
Conclusion
}
\par
\noindent

In this paper, I assert:
\begin{itemize}
\item[(G)] 
Although the von Neumann-L\"{u}ders projection postulate
(D$_2$)
concerning the measurement
${\mathsf{M}}_{B(H)} ({\mathsf{O}}_P,$
$ S_{[\rho]})$
can not be derived from Axioms 1 and 2,
the similar result (F$_2$) concerning 
${\mathsf M}_{B(H)} (\Psi ( {\mathsf O}_G \otimes {\mathsf O}_F), S_{[\rho ]} )$ holds in the linguistic interpretation.
\end{itemize}
Hence, I assert that the (D$_2$) (i.e., the wave function collapse ) should not be adopted in the Copenhagen interpretation.
Although there are a lot of opinions about the Copenhagen interpretation ({\it cf.} \cite{Howard}),
I want to conclude, as mentioned in \cite{KSTS}, that
the linguistic interpretation is the true colors of the Copenhagen interpretation.
Also, if this is true, other interpretations (e.g., the many-worlds, etc.) should be reconsidered.
\par
I hope that my assertion will be examined from various points of view.

\par \noindent
\vskip0.5cm
\par \noindent
\textcolor{red}{\large Additional information (January 9, 2016):}
\normalsize
\par
\noindent
\rm
Submission history: [v1] Fri, 30 Oct 2015 09:36:21 GMT
\par
\noindent
\rm
This preprint [v1] was published in the additional references \cite{KSTS2, JQIS3, KSTS3}.
These are easier to understand than this preprint.

\rm
\par
\renewcommand{\refname}{
\large 
References}
{
\small

\normalsize
}

\end{document}